\documentclass[prb,amsmath,amssymb]{revtex4}

\usepackage{graphicx}% Include figure files
\usepackage{dcolumn}% Align table columns on decimal point
\usepackage{bm}% bold math

\newcommand{\vecr}{\vec{r}}
\newcommand{\vecLR}{\vec{R}}

\newcommand{\vb}{\vec{b}}
\newcommand{\vk}{\vec{k}}
\newcommand{\vkB}{k_{\mbox{\scriptsize B}}}
\newcommand{\Tc}{T_{\mbox{\scriptsize c}}}
\newcommand{\gammaa}{\gamma_{\rm rc}}

\newcommand{\vecj}{\vec{j}}

\newcommand{\psitrap}{\psi_{\rm out}}
\newcommand{\hatpsitrap}{\hat{\psi}_{\rm out}}
\newcommand{\psireact}{\psi_{\rm rc}}
\newcommand{\hatpsireact}{\hat{\psi}_{\rm rc}}

\begin{document}
%\preprint{Submitted to }
%\thispagestyle{empty}
%%%%%%%%%%%%%%%%%%%%%%
%\draft
%\preprint{submitted to ...}
%%%%%%%%%%%%%%%%%%%%%%%%%%%%%%%%%%%%%%%%%%%%%%%%%%%%%%%%%%%%%%%%%%%%%%
% TITLE  %%%%%%%%%%%%%%%%%%%%%%%%%%%%%%%%%%%%%%%%%%%%%%%%%%%%%%%%%%%%%
%%%%%%%%%%%%%%%%%%%%%%%%%%%%%%%%%%%%%%%%%%%%%%%%%%%%%%%%%%%%%%%%%%%%%%
%
\title{
Fractional Reaction-Diffusion Equation
}
\author{
Kazuhiko Seki}
\email{k-seki@aist.go.jp}
\author{Mariusz Wojcik}
\altaffiliation[Permanent address:]{
Institute of Applied Radiation Chemistry, Technical University of Lodz,
Wroblewskiego 15, 93-590 Lodz, Poland
}
\email{
wojcikm@mitr.p.lodz.pl
}
\author{M. Tachiya
}
\email{m.tachiya@aist.go.jp}
\affiliation{
National Institute of Advanced Industrial Science and Technology (AIST)\\
AIST Tsukuba Central 5, Higashi 1-1-1, Tsukuba, Ibaraki, Japan, 305-8565
}
\date{\today}
%%%%%%%%%%%%%%%%%%%%%%%%%%%%%%%%%%%%%%%%%%%%%%%%%%%%%%%%%%%%%%%%%%%%%%
% ABSTRACT  %%%%%%%%%%%%%%%%%%%%%%%%%%%%%%%%%%%%%%%%%%%%%%%%%%%%%%%%%%
%%%%%%%%%%%%%%%%%%%%%%%%%%%%%%%%%%%%%%%%%%%%%%%%%%%%%%%%%%%%%%%%%%%%%%
%\thispagestyle{empty}
\begin{abstract}
%\begin{center}
%Abstract
%\end{center}
A fractional reaction-diffusion equation is derived 
from a continuous time random walk model 
when the transport is dispersive. 
The exit from the encounter distance, which is described by the algebraic
waiting time distribution of jump motion, interferes with the reaction at
the encounter distance.
Therefore, 
the reaction term has a memory effect.
The derived equation is applied to the geminate recombination problem. 
The recombination is shown to depend on the intrinsic reaction rate,  
in contrast with the results of Sung et al. 
[J. Chem. Phys. {\bf 116}, 2338 (2002)], 
which were obtained 
from the fractional reaction-diffusion equation 
where the diffusion term has a memory effect but the reaction term does not. 
The reactivity dependence of the recombination probability 
is confirmed by numerical simulations.
\end{abstract}
%\pacs{82.40.-g,02.50.Ey,05.40.+j}% PACS, the Physics and Astronomy
\maketitle

%\newpage
%%%%%%%%%%%%%%%%%%%%%%%%%%%%%%%%%%%%%%%%%%%%%%%%%%%%%%%%%%%%%%%%%%%%%%
% 1 Introduction %%%%%%%%%%%%%%%%%%%%%%%%%%%%%%%%%%%%%%%%%%%%%%%%%%%%%
%%%%%%%%%%%%%%%%%%%%%%%%%%%%%%%%%%%%%%%%%%%%%%%%%%%%%%%%%%%%%%%%%%%%%%
\setcounter{equation}{0}
\section{Introduction}
\vspace{0.5cm}

Anomalous diffusion processes occur in many physical systems for various reasons 
including disorder in terms of energy or space or both \cite{Sokolov,Hughes}.
The anomalous diffusion processes are expressed as
$\langle r^2 (t) \rangle \propto t^\alpha $ for the displacement $r(t)$; 
the process is called sub-diffusive or dispersive if $0 < \alpha < 1$ while it is 
called super-diffusive  if $\alpha > 1$ \cite{Sokolov,Hughes}.
One of the useful theories for such processes 
is the continuous time random walk where 
the non-Poissonian waiting time distribution of jump motion is introduced \cite{Montroll65}. 
For sub-diffusion processes 
the fractional diffusion equation has been derived 
from the continuous time random walk model with 
the power-law waiting time distribution function \cite{Kenkre}.
The theory is continuous in space and thus is very useful to introduce the effect of interactions and 
the boundary conditions.

The fractional diffusion equation is valuable for describing reactions 
in the dispersive transport media \cite{Yuste,Sung}. 
In the theory of diffusion-controlled reactions the boundary conditions describing reactions are 
very important.
For the perfectly absorbing boundary conditions 
the generalization of the Smoluchowski theory based on the ordinary diffusion equation 
to the fractional diffusion equation is straightforward.
On the other hand, for the partially reflecting boundary conditions the
exit from the encounter distance 
interferes with the reaction at the encounter distance, 
therefore the latter is influenced by 
the non-exponential waiting time distribution function. 
Thus, the generalization of the conventional partially reflecting boundary conditions associated with 
the ordinary diffusion equation
to those associated with fractional diffusion equation is not obvious. 
Recently, Sung et al. phenomenologically introduced the fractional reaction-diffusion equation 
and found that the recombination probability of a particle starting from $r_0$  
for the partially 
reflecting 
boundary condition is equal to 
that obtained for the perfectly absorbing boundary condition \cite{Sung}, 
and thus independent of the intrinsic reaction rate at the encounter distance. 
When the partially reflecting boundary condition is used, 
a fraction of particles that arrive at the encounter distance will recombine, 
while the others will escape the reaction.
Therefore, 
the recombination probability for the partially reflecting boundary conditions 
should be lower than that for the perfectly absorbing boundary condition. 
Fractional reaction-diffusion equations or 
continuous time random walk models are also introduced 
for the description of 
nonlinear reactions, propagating fronts and two species reactions in sub-diffusive transport media 
\cite{Henry}.
However, 
these fractional reaction-diffusion equations 
share the same structure as 
that used by Sung et al. ; 
the reaction term does not have any delay  effect and only the diffusion term has
a memory effect.
In this paper 
we derive a
fractional reaction-diffusion equation from 
a continuous time random walk model ; 
it gives the proper partially reflecting boundary conditions applicable for the fractional diffusion equation.
The reaction term as well as the diffusion term has a memory effect 
because the exit from the encounter distance, 
which is described by the algebraic waiting time distribution of jump motion 
interferes with the reaction at the encounter distance. 
Our theory is different from the more macroscopic description of nonlinear reaction 
in sub-diffusive transport media \cite{Henry}, 
where the reaction is accounted for 
not by a space dependent sink term but 
by the law of mass action  
which is space independent. 
The recombination probability is obtained from the fractional reaction-diffusion equation thus derived.
It is also shown by simulations that the recombination probability 
indeed depends on the intrinsic reaction rate.\\

%\newpage
%%%%%%%%%%%%%%%%%%%%%%%%%%%%%%%%%%%%%%%%%%%%%%%%%%%%%%%%%%%%%%%%%%%%%%
% 1 Fractional reaction-diffusion equation %%%%%%%%%%%%%%%%%%%%%%%%%%%%%%%%%%%%%%%%%%%%%%%
%%%%%%%%%%%%%%%%%%%%%%%%%%%%%%%%%%%%%%%%%%%%%%%%%%%%%%%%%%%%%%%%%%%%%%
\setcounter{equation}{0}
\section{Fractional reaction-diffusion equation}
\vspace{0.5cm}
We consider geminate recombination of a B particle 
starting at $\vec{r}_0$ 
with A.
B particle migrates toward A by anomalously slow diffusion, 
$\langle r^2 (t) \rangle \propto t^\alpha$ with $\alpha <1$.
Reaction takes place 
at the encounter distance of $R$ to $R + d r$.
After B particle leaves the encounter distance it performs  
random walk described by the sub-diffusion kinetics.
When the migration dynamics is described by the ordinary diffusion equation,
the reaction is accounted for by assuming that the flux of B into 
the region $R$ to $R+d r$ is proportional to the density of B at $R$
and the resultant boundary condition is well
established \cite{Rice}.
This boundary condition 
is referred to as the partially reflecting or radiation boundary condition.
It is also well known that imposing the boundary condition 
is equivalent to introducing the sink term in 
the diffusion equation (reaction-diffusion equation)  
with the perfectly reflecting boundary condition \cite{Rice}.
However, if the motion of B particle is sub-diffusive, 
the corresponding boundary condition is not known, 
though the sub-diffusion motion itself is described by the fractional diffusion equation.
On the other hand,  in the lattice model the sub-diffusion motion is derived from 
the theory of continuous time random walk and reaction can be 
easily accounted for in this theory.
We start from the lattice model and take the continuous limit in order to obtain the sink term for 
the fractional diffusion equation.

The continuous time random walk is specified by the waiting time distribution, 
$\psi (t)$, 
 of making a jump to a neighboring site 
at time $t$
in the absence of reaction.
To be more specific a particular model is introduced
where the algebraic asymptotic tails result from the distribution of the site energy \cite{Scher}.
The power-law waiting time distribution is also obtained from 
an exponential distribution of inter-trap distances together with 
an exponential dependence of the jump rate on the inter-trap distance \cite{Tachiya75}. 
As long as the long time behavior is concerned 
our conclusion is model-independent.
One of the site energy distribution functions 
which we encounter most frequently is the exponential distribution \cite{Scher}, 
\begin{eqnarray}
g(E)= \exp \left[ - E/(\vkB \Tc) \right] /(\vkB \Tc) .
\label{traped}
\end{eqnarray}
For the activated release rate, 
\begin{eqnarray}
\gamma (E) = \gamma_{\rm r} \exp \left[ - E/(\vkB T) \right] ,
\label{releaser}
\end{eqnarray}
the waiting time distribution for release is given by  \cite{Scher}
\begin{eqnarray}
\psi (t) = \int_0^{\infty} d\,E g(E) \gamma (E) \exp \left( - \gamma (E) t \right)
\sim \frac{\alpha \Gamma \left( \alpha + 1\right)}{\gamma_{\rm r}^\alpha t^{\alpha+1}} , 
\end{eqnarray}
where $\alpha \equiv T/\Tc$ and $\Gamma (z)$ is the Gamma function. 
Since we are interested in dispersive transport $\alpha <1$ is assumed.
In the small $s$ limit the Laplace transform 
of the waiting time distribution function can be expressed as \cite{Scher}
$
\hat{\psi} (s) 
\sim 1 - 
\left[ \pi \alpha/\sin (\pi \alpha) \right] \left( s/ \gamma_{\rm r} \right)^{\alpha} 
\mbox{ for }  s/\gamma_{\rm r} < 1 .
$
Two types of waiting time distribution are defined at the encounter distance ; 
one is the waiting time distribution function of making a jump to a neighboring lattice site $\psitrap (t)$ 
and the other is the waiting time distribution function of reaction $\psireact (t)$.
The waiting time distribution of making a jump at the encounter distance is given by, 
\begin{eqnarray}
\psitrap (t) =  \psi (t) \exp \left( - \gammaa t \right) .
\end{eqnarray}
The waiting time distribution of reaction is defined as 
the probability that the particle which is initially at a site in 
the encounter distance will undergo reaction without making a jump at time $t$. 
It is given by the 
reaction rate constant,  $\gammaa$, multiplied by the 
remaining probability of particles at the site in the reaction zone, which decays 
either by jump motion or reaction,  
\begin{eqnarray}
\psireact (t) &=& \gammaa \exp \left( - \gammaa t \right)
\int_t^\infty d\,t_1 \psi(t_1) .
\label{psireaction}
\end{eqnarray}
It is implicitly assumed in our model that 
in the presence of site energy distribution $g(E)$
the reaction takes place from any energy level with the same rate 
$\gammaa$.

We define the vector characterizing a jump to the nearest neighbor site $j$,  
by $\vb_j$ ($j=1, 2, \cdots, 2d$)
and the jump length $b$.
In the limit of small $\Delta {\cal R}$ 
the region from $R-\Delta {\cal R}/2$ to $R+\Delta {\cal R}/2$ can be regarded as the encounter distance. 
In this section we first consider the system without the reflecting boundary condition and 
introduce later the reflective sphere of radius $r=R$.
The equation governing the probability $\eta (\vecr_i,t)$ of 
just arriving at site $\vecr_i$ at time $t$ 
can be written as,
%\begin{widetext}
\begin{eqnarray}
\lefteqn{
\eta (\vecr_i, t) 
= \frac{1
}{2d}  \sum_{j=1}^{2d}
\int_0^t dt_1 \left[ \psi (t-t_1) \eta (\vecr_i -\vb_j, t_1) - \right.
} 
\nonumber  \\
& &\mbox{} 
\left.
\sum_{k} 
%\frac{\sum_{j=1}^{2d}}{2d} 
\delta_{\vecr_i -\vb_j, \vecr_k} 
%\int_0^t dt_1 
\delta \psi (t-t_1) 
\Delta H (R-\Delta {\cal R}/2-r_k; \Delta {\cal R}) 
\eta (\vecr_k, t_1) \right] 
\nonumber \\ 
& &\mbox{}
+ \delta_{\vecr_i, \vecr_0} \delta (t) 
 ,
\label{basic1}
\end{eqnarray}
%\end{widetext}
where 
$\Delta H (R-r ; {\Delta {\cal R}}) \equiv
\left[ H(R+\Delta {\cal R}-r) - H (R-r) \right]
$
and 
$H(r)$ is the Heaviside step function, namely, 
$H(r)=1$ for $r \geq 0$, otherwise $H(r)=0$.
$\delta \psi (t) \equiv \psi (t) -\psitrap (t)$ is the difference 
between the waiting time distribution of making a jump 
at other sites and that 
at the encounter distance. 
The difference arises because of the reaction at the encounter distance.
The probability of just leaving site $\vecr_i$ at time $t$ 
is given by $\int_0^t d\, t_1 \psi (t-t_1) \eta (\vecr_i, t_1)$.
By subtracting this quantity from both sides of Eq. (\ref{basic1}) 
we obtain the balance equation.
After introducing the Laplace transform, 
$\hat{\psi}(s) = \int_0^\infty d\, t \exp (- st) \psi (t)$, 
the balance equation is written as,
%\begin{widetext}
\begin{eqnarray}
\lefteqn{
\left[1-\hat{\psi} (s) \right] \hat{\eta} (\vecr_i,s) = 
\hat{\psi} (s)  \left[ \frac{1}{2d} \sum_{j=1}^{2d}
\hat{\eta} (\vecr_i -\vb_j, s) - \hat{\eta} (\vecr_i) \right] - 
}\nonumber  \\
& &
\delta \hat{\psi} (s) \frac{1}{2d} \sum_{j=1}^{2d}
\Delta H  \left( R-\Delta {\cal R}/2- \left| \vecr_i  -\vb_j \right| ; {\Delta {\cal R}} \right) 
\hat{\eta} (\vecr_i -\vb_j, s)
+ \delta_{\vecr_i, \vecr_0} .  \nonumber 
\label{basic2}
\end{eqnarray}
%\end{widetext}
Now we introduce the perfectly reflective sphere of radius $R$. 
Reaction takes place when a particle enters the reaction zone, which is 
defined by the volume between 
$R$ and $R+ \Delta {\cal R}/2$. 
The shortest distance between a point outside the reaction zone and 
the reflective sphere is $\Delta {\cal R}/2$.  
Therefore any particle which makes a jump from outside the reaction zone 
and is reflected by the sphere cannot reach  
a point outside the reaction zone, 
unless the jump length is larger than $\Delta {\cal R}$. 
Thus, we set $\Delta {\cal R}=b$ to guarantee that any particle 
which makes a jump from outside through the boundary of the reaction zone 
lands within the reaction zone. 
As we will show later by simulations, 
this choice of the reaction zone also satisfies  
the condition that practically all the particles which make a jump  from inside 
the reaction zone land outside it.
Therefore, 
the particle in the reaction zone comes from outside it, 
\begin{eqnarray}
\left[1-\hat{\psi} (s) \right] \hat{\eta} (\vecr_i,s) 
= 
\hat{\psi} (s)  \left[ \frac{1}{2d} \sum_{j=1}^{2d}
\hat{\eta} (\overline{\vecr_i -\vb_j}, s) - \hat{\eta} (\vecr_i) \right]  
+ \delta_{\vecr_i, \vecr_0}   
\mbox{ for } R \leq r_i < R + \frac{b}{2}, 
\label{basicin}
\end{eqnarray}
where $\overline{\vecr_i -\vb_j}$ 
stands for the position 
which leads to $\vecr_i$ after making a jump $\vb_j$. 
If $|\vecr_i -\vb_j| \geq R$, 
the particle is not reflected by the sphere of radius $R$, 
therefore 
$\overline{\vecr_i -\vb_j}$ 
equals to 
$\vecr_i -\vb_j$ when $|\vecr_i -\vb_j| \geq R$. 
On the other hand,  
if $|\vecr_i -\vb_j| < R$, 
the particle is reflected by the sphere. 
In this case  
$\overline{\vecr_i -\vb_j}$ 
is given by the position shown in 
Fig. 1 a) and b).
Strictly speaking, 
even if $|\vecr_i -\vb_j| > R$, 
there is a case where a particle is actually reflected by the sphere, 
as shown in Fig. 1 c).
In this case 
$\overline{\vecr_i -\vb_j} \neq \vecr_i - \vb_j$.  
But such events are obviously rare. 
The balance equation outside the reaction zone obeys, 
\begin{eqnarray}
\lefteqn{
\left[1-\hat{\psi} (s) \right] \hat{\eta} (\vecr_i,s) 
= 
\hat{\psi} (s)  \left[ \frac{1}{2d} \sum_{j=1}^{2d}
\hat{\eta} (\overline{\vecr_i -\vb_j}, s) - \hat{\eta} (\vecr_i) \right] - 
}\nonumber  \\
& &
\delta \hat{\psi} (s) \frac{1}{2d} \sum_{j=1}^{2d}
\Delta H \left( R - \left| \overline{\vecr_i  -\vb_j} \right|; \frac{b}{2} \right) 
\hat{\eta} (\overline{\vecr_i -\vb_j}, s)
+ \delta_{\vecr_i, \vecr_0} 
~~~~~~\mbox{ for } R + \frac{b}{2} \leq r_i .   
\label{basicout}
\end{eqnarray}
We introduce the corresponding probability density,
$
\eta \left( \vecr, t \right) = \sum_i \eta \left( \vecr_i, t \right) \delta \left( \vecr - \vecr_i \right) 
$. 
Eq. (\ref{basicin}) does not directly involve the effect of reaction, 
while Eq. (\ref{basicout}) does.
Moreover, 
for a small jump length 
the contribution from 
Eq. (\ref{basicin}) is negligible. 
Therefore,  
we focus our attention on Eq. (\ref{basicout}).
Performing Taylor expansion,
the leading terms satisfy the differential equation, 
%\begin{widetext}
\begin{eqnarray}
\lefteqn{\left[1-\hat{\psi} (s) \right] \hat{\eta} (\vecr,s) =
 \frac{b^2}{2} \hat{\psi}(s) \nabla^2 \hat{\eta} \left( \vecr, s \right) - } \nonumber \\
&&
\frac{b}{2} \delta \hat{\psi} (s) \delta \left( R -r \right) \hat{\eta} \left( \vecr, s \right) +
\delta \left( \vecr - \vecr_0 \right) ,
\label{diffeeq}
\end{eqnarray}
%\end{widetext}
where the following approximation is employed,
%\begin{widetext}
\begin{eqnarray}
\lefteqn{\frac{1}{2d} \sum_{j=1}^{2d} 
\Delta H \left( R- \left| \overline{\vecr  -\vb_j} \right| ;\frac{b}{2} \right)
=
\frac{1}{2d} \sum_{j=1}^{2d} 
\Delta H \left( R-b/2- \left| \vecr  -\vb_j \right| ; b \right)  
}\nonumber \\
&&
= \frac{1}{2}
\Delta H \left( R  +b/2 - r ; b \right)
= \frac{b}{2} \delta (R+b/2-r) \sim  \frac{b}{2} \delta (R-r)  .
\label{approx}
\end{eqnarray}
%\end{widetext}
The first equality follows from  
the definition of $\overline{\vecr  -\vb_j}$, 
which is equal to $\vecr  -\vb_j$ when $R \leq \left| \vecr  -\vb_j \right| < R +b/2$, 
while it is given by the corresponding reflected vector when 
$R-b/2 \leq\left| \vecr  -\vb_j \right| < R$ as described in Fig. 1 b).
The quantity in the summation becomes $\Delta H \left( R  +b/2 - r ;b \right)$ 
for $\vb_j$ directed outward from the sphere of radius $R$ and vanishes otherwise,  
and among $2d$ of $\vb_j$ vectors half of them are directed outward. 
This explains the second equality of Eq. (\ref{approx}).
The third equality is due to an approximate expression of 
the delta function given by the first derivative of the  Heaviside step function.
\begin{figure}[h]
\includegraphics[width=10cm]{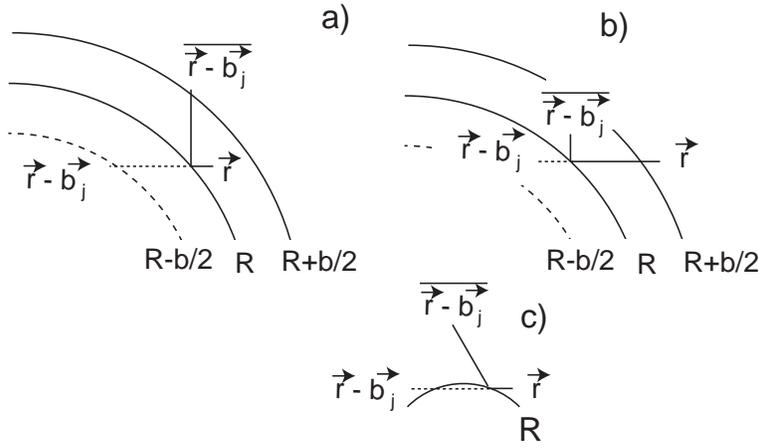}
\caption{\label{fig:reflect}
The position $\overline{\vecr -\vb_j}$ which leads to $\vecr_i$ after making a jump $\vb_j$. 
a) for $|\vecr -\vb_j| < R$ and $R < r <R+b/2$.
b) for $|\vecr -\vb_j| < R$ and $R +b/2 < r $.
c) for  $|\vecr -\vb_j| > R$. 
In this case, 
although $|\vecr -\vb_j| > R$, 
the particle is reflected and 
$\overline{\vecr -\vb_j}$ is actually not equal to $\vecr -\vb_j$.
}
\end{figure}
So far, 
we have considered the probability density of just arriving 
at $\vecr$ 
at time $t$
, $\eta (\vecr , t)$.
The usual probability density is defined in terms of 
the remaining probability,
$\phi (t) \equiv \int_t^\infty d\, t_1 \psi (t_1) $ as 
$\rho \left( \vecr, t \right) = \int_0^t d\, t_1 \phi \left( t -t_1\right) 
\eta \left( \vecr ,t_1 \right).
$
By noticing the relation, $\hat{\phi} (s) = \left( 1 - \hat{\psi} (s) \right)/s$, 
Eq. (\ref{diffeeq}) becomes
%\begin{widetext}
\begin{eqnarray}
\lefteqn{ 
s \hat{\rho} (\vecr,s) - \rho \left( \vecr, t=0 \right) = 
\frac{s}{1 - \hat{\psi} (s)} 
 \times
} 
\nonumber \\
&&
\left( \frac{b^2}{2} \hat{\psi} (s) \nabla^2 \hat{\rho} (\vecr,s)- 2 \pi R^2 b \delta \hat{\psi} (s) 
\frac{\delta (r-R) }{4 \pi R^2} \hat{\rho} (\vecr,s) \right)  .
\label{fundamental}
\end{eqnarray}
Since a particle cannot penetrate the sphere of radius $R$, 
the perfectly reflecting boundary is imposed at $r=R$. 
In the long time limit which is expressed in the Laplace domain as $s \rightarrow 0$ 
we have $1- \hat{\psi} (s) \rightarrow s^{\alpha}$ and Eq. (\ref{fundamental}) reduces to
%\begin{widetext}
\begin{eqnarray}
\lefteqn{
s \hat{\rho} (\vecr,s) - \rho \left( \vecr, t=0 \right) =
\frac{s}{1 - \hat{\psi} (s)} 
\times } \nonumber \\
&&
\left( \frac{b^2}{2} \nabla^2 \hat{\rho} (\vecr,s)- 2 \pi R^2 b \hatpsireact (0) 
\frac{\delta (r-R) }{4 \pi R^2} \hat{\rho} (\vecr,s) \right)  .
\label{fundamentala}
\end{eqnarray}
%\end{widetext}
where $\hatpsireact (0) = 1 - \hatpsitrap (0)$ 
is introduced,  
which follows from the fact that particles at a given site in the reaction zone 
perform either jump or reaction, 
namely 
$\int_0^\infty \psireact (t) d\,t + \int_0^\infty \psitrap (t) d\, t=1$ ; 
the overall probability of reaction,  
$\hatpsireact (0)$, 
is the branching ratio of undergoing reaction at a site in the reaction zone and 
$\hatpsitrap (0)$ represents 
the branching ratio of making a jump at the same site. 
For the normal diffusion, $\alpha=1$, 
Eq. (\ref{fundamentala}) yields the usual reaction-diffusion equation 
after inverse Laplace transform. 
For sub-diffusive transport media, $\alpha < 1$, 
the right-hand side of Eq. (\ref{fundamentala})  
is given in the time domain by 
time differentiation of 
the convolution of the retardation function 
of algebraic form 
with the function which includes  
both the diffusion term  
and the reaction term.
Now, we investigate this feature.

The Laplace transform of $\psireact(t) $ defined in Eq. (\ref{psireaction})  is obtained as 
$
\hatpsireact (s) \sim 1 - \hat{\psi} \left(s + \gammaa \right)  .
$
In the limit of fast release rate $\gamma_{\rm r}$ and fast reaction rate $\gammaa$ 
we get
\begin{eqnarray}
\hatpsireact (s) \sim 
\frac{\pi \alpha}{\sin \pi \alpha} \left( \frac{\gammaa}{\gamma_{\rm r}} \right)^{\alpha} 
\label{scaling}
\end{eqnarray}
for reaction-limited condition, $\gammaa < \gamma_{\rm r}$. 
In this case $\hatpsireact (s)$ is a constant independent of $s$.
On the basis of the diffusion-reaction model described above we define 
the generalized diffusion constant $D_\alpha$ as 
\begin{eqnarray} 
D_\alpha \equiv \frac{\sin \pi \alpha}{2 \pi \alpha} \gamma_{\rm r}^{\alpha} b^2 .
\label{defD}
\end{eqnarray}
Then the 
inverse Laplace transform of Eq. (\ref{fundamentala}) 
is obtained as 
\begin{eqnarray}
\lefteqn{
\frac{\partial}{\partial t} \rho \left( r, t \right) = \frac{\partial}{\partial t}
\int_0^t d\,t_1 \frac{1}{\Gamma ( \alpha )} \frac{1}{\left( t - t_1 \right)^{1-\alpha}}
\times} \nonumber \\ & &
\left[ D_{\alpha} \nabla^2 \rho \left( r, t_1 \right) - k_{\alpha} 
\frac{\delta \left( r - R \right)}{4 \pi R^2} \rho \left( r, t_1 \right) \right] ,
\label{fundamentale}
\end{eqnarray}
where the generalized intrinsic reaction rate $k_\alpha$ is defined as ,
\begin{eqnarray} 
k_{\alpha} \equiv \gammaa^\alpha 2 \pi R^2 b .
\label{defk}
\end{eqnarray}
The exit from the encounter distance, 
which is described by the algebraic waiting time distribution of jump motion 
interferes with the reaction at the encounter distance. 
Therefore, the reaction term has a memory effect like the term describing sub-diffusion motion.
So far $b$ has been assumed to be small but finite and 
we have investigated equations under the condition of large $\gamma_{\rm r}$ and $\gammaa$.
In order to derive the equations 
valid in the continuous limit 
we have to take the limit of 
$b \rightarrow 0$,  
$\gamma_{\rm r} \rightarrow \infty$ and $\gammaa \rightarrow \infty$ 
with 
$D_\alpha$ and $k_\alpha$ kept finite.
Here, 
the fractional reaction-diffusion equation has been derived 
for a specific initial condition, 
but the linearity of the model guarantees a much wider range of applicability:
essentially, any initial distribution will be acceptable.\\

In the reaction-diffusion equation, Eq. (\ref{fundamentale}),  
the reaction is accounted for by introducing the sink term, 
with the perfectly reflecting boundary condition imposed.
If the reaction is accounted for by imposing 
the boundary condition associated with the 
fractional diffusion equation \cite{Kenkre},
\begin{eqnarray}
\frac{\partial}{\partial t} \rho \left( r, t \right) = \frac{\partial}{\partial t}
\int_0^t d\,t_1 \frac{1}{\Gamma ( \alpha )} \frac{1}{\left( t - t_1 \right)^{1-\alpha}}
D_{\alpha} \nabla^2 \rho \left( r, t_1 \right) ,
\label{fd}
\end{eqnarray}
what is the relevant boundary condition? 
By multiplying both sides of Eq. (\ref{fundamentale}) 
by $\int_R^{R+\epsilon} 4 \pi r^2 d\, r$ 
we obtain in the limit of a small $\epsilon$,
\begin{eqnarray}
\left. D_\alpha \frac{\partial}{\partial r} \rho \left(r, t \right) \right|_{r=R+\epsilon}
= \frac{k_{\alpha}}{4 \pi R^2} \rho \left(R, t \right) .
\label{boundary}
\end{eqnarray}
Therefore 
the boundary condition for the fractional diffusion equation, Eq. (\ref{fd}), is the 
$\epsilon \rightarrow 0$ limit of  Eq. (\ref{boundary}).\\
Eq. (\ref{fundamentale}) or Eq.(\ref{fd}) with Eq. (\ref{boundary}) is the most important result of 
this paper.

%\newpage
%%%%%%%%%%%%%%%%%%%%%%%%%%%%%%%%%%%%%%%%%%%%%%%%%%%%%%%%%%%%%%%%%%%%%%
% Recombination probability%%%%%%%%%%%%%%%%%%%%%%%%%%%%%%%%%%%%%%%%%%%%%%%
%%%%%%%%%%%%%%%%%%%%%%%%%%%%%%%%%%%%%%%%%%%%%%%%%%%%%%%%%%%%%%%%%%%%%%
\setcounter{equation}{0}
\section{Recombination probability}
\vspace{0.5cm}

Now we apply our equation to the geminate recombination problem. 
The recombination probability is obtained by the usual procedure.
The fractional diffusion equation can be cast into the form,
$
 \frac{\partial}{\partial t} \rho \left( r, t \right) = - \mbox{div} \vecj (r, t) ,
$
where the rate of the change of the density is related to the current defined by,
$
\vecj (r, t) \equiv - \frac{\partial}{\partial t}
\int_0^t d\,t_1 \frac{1}{\Gamma ( \alpha )} \frac{1}{\left( t - t_1 \right)^{1-\alpha}}
D_{\alpha} \mbox{grad} \rho \left( r, t_1 \right) .
$
The recombination probability $\kappa \left( r_0 \right)$ is expressed as 
$
\kappa \left( r_0 \right) =- \int_0^\infty d\, t_1 4 \pi R^2 \vecj (R, t_1) \cdot \vecLR/R ,
$
which can be calculated by the standard method as,
\begin{eqnarray}
\kappa \left( r_0 \right) = \frac{R}{r_0} \frac{1}{\displaystyle 1 + 
\frac{4 \pi R D_{\alpha}}{k_\alpha}} .
\label{result}
\end{eqnarray}
The familiar form of the recombination probability is derived, 
which clearly shows its dependence on the intrinsic reaction rate, 
in contrast with the results of Sung et al. \cite{Sung}, 
which were obtained from  
the fractional reaction-diffusion equation 
where only the diffusion term has 
a memory effect  and the 
reaction term does not. 
One should also note that both the diffusion constant and the intrinsic reaction rate 
scale with parameter $\alpha$  as shown in Eqs. (\ref{defD}) and (\ref{defk}).
The scaling for the intrinsic reaction rate is due to the reaction model used.
For other models 
different scaling with $\alpha$ may be derived.\\

Eq. (\ref{result}) can be derived more directly in the following way.
Since the probability that a particle which starts at $r_0$ 
will visit the spherical shell of radius $R$ 
is  equal to 
the recombination probability 
for the perfectly absorbing boundary condition \cite{Rice}, 
it is given by  
$
R/r_0 
$.
Inside the reaction zone the probability that a particle makes a jump without 
reaction is $\hatpsitrap (0)$. 
After a jump the particles in the reaction zone may still remain inside the reaction zone or 
leave the reaction zone. 
In the limit that the width of the reaction zone goes to zero, 
we consider two probabilities, namely,
the probability that a particle at 
the spherical shell of radius $R$ will make a jump 
to that of radius $R+b$ with $b$ being the jump length,
$
m \hatpsitrap (0) ,
$
and that of making a jump to another position at the 
reaction distance $R$, 
$
n\hatpsitrap (0) 
$, 
where 
$m$ and $n$ are the branching ratios of respective jumps 
and 
$m+n=1$.
On the other hand, the probability 
that it
will undergo reaction is
$
\hatpsireact (0) .
$
Some particles may recombine at the first visit to the encounter distance $R$.
Other part of particles escape recombination 
at the first encounter and make a jump to the sphere of 
radius $R+b$ or make a jump to another position at the encounter distance.
In the former case some particles may recombine at the second encounter after jumping from 
the sphere of radius $R+b$ to the encounter distance
or again escape recombination at the second encounter. 
In the latter case 
particles may recombine at the same position or make another jump, and so forth.  
Accordingly, 
the probability that a particle 
which starts at $r_0$ will ultimately undergo reaction is,
\begin{eqnarray}
\kappa \left( r_0 \right) &=& \hatpsireact (0) \frac{R}{r_0}
+ \hatpsireact (0)
 \left( \frac{R}{R+b}  m \hatpsitrap (0) + n \hatpsitrap (0) \right) 
 \frac{R}{r_0} + \cdots \nonumber \\
&=& \frac{R}{r_0}
 \frac{\displaystyle  \hatpsireact (0)}{\displaystyle 1 
-\left(m \hatpsitrap (0) \frac{R}{R+b} + n \hatpsitrap (0) \right)} \nonumber \\
&=& \frac{R}{r_0} \frac{1}{\displaystyle 
\frac{R+n b}{R+b} + \frac{m b}{R+b} 
\frac{1}{\displaystyle \hatpsireact (0)} } \nonumber \\
&\sim & \frac{R}{r_0} \frac{1}{\displaystyle 
1 + \frac{m b}{R} 
\frac{1}{\displaystyle \hatpsireact (0)} } .
\label{TachiyaRBP}
\end{eqnarray}
Therefore, 
the recombination probability is given by  
Eq. (\ref{result}) 
with $k_\alpha$ generalized to,
\begin{eqnarray}
k_{\alpha} \sim 2 \pi R^2 b \gamma_{\rm rc}^\alpha /m  .
\label{generalizedk}
\end{eqnarray}
In simulations we cannot take the zero limit for the width of the reaction zone 
and the value of $m$ 
depends on the width. 
By comparing to the simulation results we find
$m=1$ for the width of $b/2$ and $m=1/2$ for the width of $b$, as will be shown later 
when we compare the analytical result with simulations. 
Eq. (\ref{fundamentale}) is derived for the width of the reaction zone equal to $b/2$ 
before taking the limit that the jump length goes to zero.
Since $m=1$ is obtained for this choice of 
the width, 
the intrinsic reaction rate Eq. (\ref{defk}) is consistent with 
the definition of 
the generalized intrinsic reaction rate, Eq. (\ref{generalizedk}). \\
%\newpage
%%%%%%%%%%%%%%%%%%%%%%%%%%%%%%%%%%%%%%%%%%%%%%%%%%%%%%%%%%%%%%%%%%%%%%
% Simulations %%%%%%%%%%%%%%%%%%%%%%%%%%%%%%%%%%%%%%%%%%%%%%%
%%%%%%%%%%%%%%%%%%%%%%%%%%%%%%%%%%%%%%%%%%%%%%%%%%%%%%%%%%%%%%%%%%%%%%
\setcounter{equation}{0}
\section{Simulations}
\vspace{0.5cm}
In order to confirm the validity of the above results 
we have carried out simulations. 
In our calculations 
the random walk is realized as a sequence of 
instantaneous jumps between the traps, 
with the trap energies generated according to the distribution
Eq. (\ref{traped}). 
The rate of release from the traps is assumed in the form given by 
Eq. (\ref{releaser}), 
and the detrapping time for a trap with energy $E$  
is obtained from the exponential distribution with the mean value 
$1/\gamma (E) $ . 
We assume 
the Gaussian distribution of jump lengths 
and calculate displacements of particle B along each Cartesian coordinate as
$
\Delta x, \Delta y, \Delta z = b \times N (0,1) ,
$
where $N (0,1)$ represents a random number 
obtained from the standard normal distribution. 
The trajectory of B is calculated 
until it either reacts with A or escapes to a large distance $r_{\rm max}$ . 
The simulation is repeated for a large number of independent trajectories, 
which allows us to calculate the reaction (and escape) probability.

We assume that the reaction occurs within a thin spherical shell 
$R - \Delta R/2 < r < R + \Delta R/2$ , where  $R$ is the reaction radius, 
and is characterized by the reaction rate  $\gamma_{\rm rc}$. 
In the simulation the reaction is modeled in the following way. 
When a particle B is trapped within the reaction shell, 
we generate the reaction time $t_{\rm rc}$  
from the exponential distribution with the mean value $1/\gamma_{\rm rc}$ , 
and compare it with the detrapping time $t$  
calculated for the current trap. 
If  $t_{\rm rc}<t$ then the reaction occurs, 
otherwise the particle jumps to another trap 
and the simulation is continued. 
A reflective boundary is established at $r=R-\Delta R/2$ , 
and the particle is bounced back when a jump across this boundary is attempted. 
The reaction model described above represents 
a partially diffusion-controlled reaction 
with the classical second-order rate constant  $k=4 \pi R^2 \Delta R \gamma_{\rm rc}$ 
when the transport is described by the normal diffusion.

The calculations presented in this work were carried out for $r_0=2R$  
and  $r_{\rm max}=100R$. 
Error in the calculated escape probability due to the finite value of  $r_{\rm max}$  
is estimated as about $1$ \%. 
The number of independent trajectories generated in each simulation 
was at least $10^4$. 
The jump length parameter was assumed as $b=0.1R$, 
and the values $\Delta R=b$ and $\Delta R = b/2$  were used. 
In the calculations we used reduced units, 
with $R$  taken as the unit of length and  $1/\gamma_{\rm r}$ as the unit of time.
The calculated diffusion constant is in good agreement with Eq. (\ref{defD}).
The simulation results also show that 
the escape probability $1 - \kappa \left( r_0 \right)$ 
depends on the intrinsic reactivity 
not only for the normal diffusion, 
but also in the sub-diffusive case as shown in Fig. 2.
\begin{figure}[h]
\includegraphics{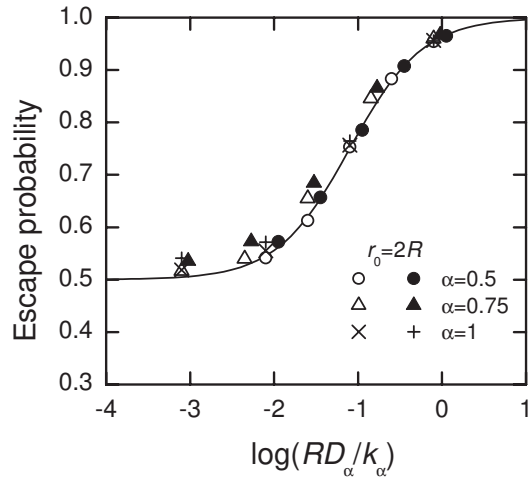}
\caption{\label{fig:escape} 
The escape probability versus $\log_{10} 
\left(R D_{\alpha}/k_\alpha \right)$.
Initial distance is $r_0=2R$.
The symbols denote simulation data. 
The closed circles, closed triangles and 
pluses correspond to the 
$\alpha$ values of $0.5$, $0.75$ and $1.0$, respectively for the width of the reaction zone $b/2$.
The open 
circles, open triangles and 
crosses correspond to the 
$\alpha$ values of $0.5$, $0.75$ and $1.0$, respectively 
for the width of the reaction zone $b$.
The line shows the analytic result of Eq. (\ref{result}).
}
\end{figure}
The analytical result of Eq. (\ref{result}) 
coincides very well with the simulation data.
When the width of the reaction zone is $b/2$ we use $m=1$ and 
when the width of the reaction zone is $b$ we use $m=1/2$ for the definition of 
the generalized intrinsic reaction rate, Eq. (\ref{generalizedk}).
Since the simulation data coincide very well with the analytical result of Eq. (\ref{result}) 
the above choice of $m$ is also justified.
When the width is the same as the jump length of the particle motion, $b$, 
the particles which make a jump from inside the reaction zone 
and are reflected by the sphere of radius $R$ 
still remain inside the reaction zone.
Approximately half of the particles make a jump outward from 
the sphere of radius $R$ and 
leave the reaction zone after a jump.
On the other hand, when the width is as small as $b/2$,  
almost all the particles including those reflected by the sphere of radius $R$ 
leave the reaction zone after a jump.\\

%\newpage
%%%%%%%%%%%%%%%%%%%%%%%%%%%%%%%%%%%%%%%%%%%%%%%%%%%%%%%%%%%%%%%%%%%%%%
% Recombination probability in a lattice model %%%%%%%%%%%%%%%%%%%%%%%%%%%%%%%%%%%%%%%%%%%%%%%
%%%%%%%%%%%%%%%%%%%%%%%%%%%%%%%%%%%%%%%%%%%%%%%%%%%%%%%%%%%%%%%%%%%%%%
\setcounter{equation}{0}
\section{Recombination probability in a lattice model}
\vspace{0.5cm}
Finally we analyze the reaction kinetics in 
the full lattice model where the reaction proceeds only at the origin.
The recombination probability of a particle starting from $\vecr_0$ 
for the partially reflecting boundary condition  
is obtained
from the known result 
for the perfectly absorbing boundary condition
by the method of Pedersen 
\cite{Pedersen,Watanabe},
\begin{eqnarray}
\kappa \left( r_0 \right) = \frac{\displaystyle \frac{1}{(2\pi)^d}\int \cdots  \int_{-\pi}^{\pi} d^d \vk
\frac{\cos (\vk \cdot \vecr_0 /b)}{1 - \lambda (\vk)}}
{\displaystyle \frac{1}{\hatpsireact (0)} +  \frac{1}{(2\pi)^d} \int \cdots \int_{-\pi}^{\pi} d^d \vk
\frac{\lambda (\vk) } 
{1 - \lambda (\vk) }}  ,
\label{RPl}
\end{eqnarray}
where $\lambda (\vk)$ is the structure factor defined by
$
\lambda (\vk) \equiv \frac{1}{2d} \sum_{j=1}^{2d} \cos \left(\vk \cdot \vb_j/b \right) .
$
For a particle starting from the first neighbor in Simple Cubic lattice \cite{Hughes},
\begin{eqnarray*}
\lefteqn{
\frac{1}{(2\pi)^d} \int \cdots \int_{-\pi}^{\pi} d^d \vk
\frac{\cos (\vk \cdot \vecr_0 /b)}{1 - \lambda (\vk)} =
} \\
&& 
 \frac{1}{(2\pi)^d} \int \cdots \int_{-\pi}^{\pi} d^d \vk
\frac{\lambda (\vk) } 
{1 - \lambda (\vk) } 
= 0.516386 . 
\end{eqnarray*}
Eq. (\ref{RPl})  together with 
Eq. (\ref{scaling}) for 
$
\hatpsireact (s) 
$
yields again the recombination probability as a function of the intrinsic reaction rate $\gammaa$ 
at the origin.\\

%\newpage
%%%%%%%%%%%%%%%%%%%%%%%%%%%%%%%%%%%%%%%%%%%%%%%%%%%%%%%%%%%%%%%%%%%%%%
% Conclusions %%%%%%%%%%%%%%%%%%%%%%%%%%%%%%%%%%%%%%%%%%%%%%%
%%%%%%%%%%%%%%%%%%%%%%%%%%%%%%%%%%%%%%%%%%%%%%%%%%%%%%%%%%%%%%%%%%%%%%
\setcounter{equation}{0}
\section{Conclusions}
\vspace{0.5cm}

In the absence of reaction 
it is well known that 
the fractional diffusion equation is derived from the continuous time 
random walk models \cite{Sokolov,Hughes,Kenkre}. 
Therefore, 
the continuous time random walk model 
can be regarded as a basis for the fractional diffusion equation 
describing sub-diffusive transport. 
In this paper 
a fractional reaction-diffusion equation is derived from 
a continuous time random walk model. 
The reaction term has a memory effect 
because the exit from the encounter distance 
which is described by the algebraic asymptotic form of 
the waiting time distribution of jump motion 
interferes with the reaction at the encounter distance.
From the fractional reaction-diffusion equation thus derived 
the recombination probability 
is obtained, 
which depends on the intrinsic reaction rate for the 
sub-diffusive case as well as for the normal diffusion, 
unlike the result of Sung et al. \cite{Sung}.
The theory of Sung et al. is based on the fractional reaction-diffusion equation where 
only the diffusion term has a memory effect and the reaction term does not \cite{Sung}. 
They obtained the recombination probability for the partially 
reflecting 
boundary condition which is equal to 
that obtained for the perfectly absorbing boundary condition \cite{Sung}. 
An argument that is sometimes used to justify their result 
is that for the sub-diffusive case 
the mean residence time at a site is infinitely long, 
therefore any particle in a reactive zone 
undergoes reaction 
however small the intrinsic reaction rate may be. 
We have shown here that even if the mean residence time is infinite, 
each particle at a site 
has a finite waiting time of jump motion and 
a fraction of particles in the reaction zone will escape reaction,  
if the intrinsic reaction rate is finite.  
This is also clear from our simulation procedure and from the 
argument leading to Eq. (\ref{TachiyaRBP}) and its counterpart in the lattice model, Eq. (\ref{RPl}).
To corroborate the above argument 
the fractional reaction-diffusion equation is derived, 
which has memory effects both in diffusion term and in reaction term. 
The analytical expression of the recombination probability has been 
derived from the fractional reaction-diffusion equation and 
coincides very well with the simulation data. 
The memory effect in the reaction term is due to 
the interference of the reaction at the encounter distance with 
the exit from the encounter distance which has dispersive kinetics 
in a sub-diffusive transport media. 
Although in the present paper the fractional reaction-diffusion equation has been derived  
for a specific initial condition, 
the linearity of the model guarantees a much wider range of applicability:
essentially, any initial distribution will be acceptable.
Finally, we remark that in our study the reaction is assumed to take place at some distance. 
Most of the theories on reactions in sub-diffusive transport are developed  
on the basis of more macroscopic description \cite{Henry}, 
where the reaction is accounted for by 
the law of mass action which is space independent. 
In those theories, 
reaction terms without memory effect are simply added to the diffusion term with 
memory kernel. 
In this paper reactions are accounted for by the waiting time distribution functions in the reaction zone 
and the memory effects in reaction term as well as in diffusion term 
automatically emerge from such description.

%%%%%%%%%%%%%%%%%%%%%%%%%%%%%%%%%%%%%%%%%%%%%%%%%%%%%%%%%%%%%%%%%%%%%%
% ACKNOWLEDGMENTS %%%%%%%%%%%%%%%%%%%%%%%%%%%%%%%%%%%%%%%%%%%%%%%%%%%%
%%%%%%%%%%%%%%%%%%%%%%%%%%%%%%%%%%%%%%%%%%%%%%%%%%%%%%%%%%%%%%%%%%%%%%
%\newpage
%\noindent{\Large\bf Acknowledgment}
%\vspace{0.5cm}
\acknowledgments

This work is supported by the COE development program of 
MEXT
and the Grant-in-Aid for Young Scientists(B) (14740243) from 
MEXT.

%%%%%%%%%%%%%%%%%%%%%%%%%%%%%%%%%%%%%%%%%%%%%%%%%%%%%%%%%%%%%%%%%%%%%%
% REFERENCES %%%%%%%%%%%%%%%%%%%%%%%%%%%%%%%%%%%%%%%%%%%%%%%%%%%%%%%%%
%%%%%%%%%%%%%%%%%%%%%%%%%%%%%%%%%%%%%%%%%%%%%%%%%%%%%%%%%%%%%%%%%%%%%%
%\newpage
%\begin{thebibliography}{99}
%\vspace{0.5cm}

\end{document}